\documentclass[namedreferences]{SolarPhysics}
\usepackage[optionalrh]{spr-sola-addons}  
\usepackage{epsfig}                     
\usepackage{graphicx}               
\usepackage{courier}                    
\usepackage{color}   
\usepackage{url}                     
\usepackage{setspace}

\newcommand{\pd}[2]{\frac{\partial #1}{\partial #2}}
\newcommand{\pdd}[2]{\frac{\partial^2 #1}{\partial #2^2}}  
                    
\newcommand{\mb}[1]{\mathbf{#1}}                           
                  
\newcommand{\ocite}[1]{\citeauthor{#1} (\citeyear{#1})}

\newcommand{\aap}{    {\it Astron. Astrophys.}}

\newcommand{\apj}{    {\it Astrophys. J.}}

\newcommand{\grl}{    {\it Geophys. Res. Lett.}}

\newcommand{\jgr}{    {\it J. Geophys. Res.}}

\newcommand{\nar}{    {\it New Astron. Rev.}}

\newcommand{\solphys}{{\it Solar Phys.}}

\newcommand{\ssr}{    {\it Space Sci. Rev.}}

\begin{document}
\begin{article}

\doublespacing

\begin{opening}
\title{A Bayesian Approach to Forecasting Solar 
        Cycles Using a Fokker--Planck Equation}
\author{P.L.~\surname{Noble} \sep
        M.S.~\surname{Wheatland}$^{1}$ \sep
        }
\runningauthor{Noble, P.~L., Wheatland, M.~S.}
\runningtitle{A Bayesian Approach to Forecasting 
              Solar Cycles Using a Fokker--Planck 
              Equation} 
\institute{$^{1}$Sydney Institute for Astronomy, \\
                   School of Physics, \\
                   The University of Sydney, \\
                   Sydney NSW 2006, \\
                   Australia \\
                   email: \url{p.noble@physics.usyd.edu.au} 
                   }

\begin{abstract} 
A Bayesian method for forecasting solar cycles is presented.  The
approach combines a Fokker--Planck description of short--timescale
(daily) fluctuations in sunspot number (\citeauthor{NobleEtAl2011}, 
2011, \apj{} \textbf{732}, 5) with
information from other sources, such as precursor and/or dynamo models.
The forecasting is illustrated in application to two historical cycles
(cycles 19 and 20), and then to the current solar cycle (cycle 24).  
The new method allows the
prediction of quantiles, {\it i.e.} the probability that the sunspot number
falls outside large or small bounds at a given future time.  It also
permits Monte Carlo simulations to identify the expected size and timing
of the peak daily sunspot number, as well as the smoothed sunspot number 
for a cycle.  These simulations show how the large variance in daily sunspot 
number determines the actual reliability of any forecast of the smoothed 
maximum of a cycle.  For cycle 24 we forecast a maximum daily sunspot 
number of $166\pm 24$, to occur in March 2013, and a maximum value of the 
smoothed sunspot number of $66\pm5$, indicating a very small solar cycle.
\end{abstract}

\end{opening}

\section{Introduction} 
The solar cycle is the semi--periodic change in the level of magnetic
activity on the Sun, driven by a cyclical change in the Sun's local
magnetic field.  The International Sunspot Number\footnote[2]{Sunspot data
are provided by the US National Geophysical Data Center (NGDC) at
http://www.ngdc.noaa.gov/stp/spaceweather.html.}, defined as
\begin{equation} s = k(10g + N), \end{equation} where $g$ is the number
of sunspot groups, $N$ is the number of individual spots, and $k$ is a
correction factor, is the most commonly used measure of solar activity
\cite{BruzekEtAl1977}. Variation in the sunspot number is comprised of
the semi--regular 11 year cycle, as well as large daily, weekly, and
yearly fluctuations on top of the underlying secular variation
\cite{AbreuEtAl2008,NobleEtAl2011}.

The sunspot number is an important indicator of solar activity, and
hence of the space weather experienced on the Earth \cite{Petrovay2010}.  As
a result reliable prediction of the sunspot number is important.  There
are several methods for forecasting sunspot numbers (see
\opencite{Petrovay2010}; \opencite{Kane2007}; \opencite{Pesnell2008} for
reviews), including precursor methods, time--series methods, and dynamo
model based methods.  Precursor methods correlate aspects of a future
solar cycle ({\it e.g.} sunspot maximum) with indices of current solar and 
geomagnetic
activity. Time--series methods extrapolate sunspot data into the future
using mathematical or statistical techniques, including nonlinear models
({\it e.g.} \opencite{AguirreEtAl2008}; \opencite{HanslmeierEtAl2010}; and
\opencite{LetellierEtAl2006}), statistical techniques ({\it e.g.}
\opencite{Akaike1978}; \opencite{Yule1927}; \opencite{AllenEtAl2010}),
and neural networks ({\it e.g.} \opencite{Conway1998}).  Physical predictions
are provided by dynamo--based models, which start with models of the
Sun's internal dynamo, the source of the magnetic fields of sunspots
({\it e.g.} \opencite{DikpatiEtAl2006a}; \opencite{DikpatiEtAl2006b}). 
Precursor, time series, and dynamo--based methods all forecast solar
activity with some success \cite{Hathaway2009}, with precursor methods
being the most successful \cite{Kane2007}.

Two specific criticisms of existing approaches to prediction are that it
is difficult to rigorously combine forecasts from the competing methods,
and that it is unclear how to update and/or reconcile forecasts (in
particular long--range precursor forecasts) with new sunspot data as it
becomes available.  These problems were addressed by
\ocite{HathawayEtAl1999}, who used precursor methods to forecast the size
of the underlying cycle and then regressions to update precursor
forecasts as new data became available.  In this paper we also
present an approach to prediction which combines precursor and
time--series methods.  Our method is similar to that of
\ocite{HathawayEtAl1999}, with some specific differences. We combine
sunspot data with precursor forecasts in a statistically rigorous way
using a Bayesian framework, rather than combining methods using weighted
averages. We also forecast the short time--scale fluctuations in sunspot
number ({\it e.g.} the variance in the daily sunspot number), in addition to the size
and shape of the underlying solar cycle.

Section \ref{section:BayesianApproach} covers the theory of the Bayesian approach
to forecasting daily sunspot numbers.  Section \ref{ssection:FPModel}
introduces the Fokker--Planck model used to describe the
distribution of sunspot numbers, and Section \ref{ssection:Bayesian}
gives the details of the Bayesian method.  Section \ref{ssection:mean}
discusses the historical average solar cycle, or `mean cycle', which we use
as a starting point for prediction.  Section 
\ref{ssection:Simulation} introduces a Monte--Carlo approach
to simulating daily sunspot numbers based on analytic approximations to the
Fokker--Planck model, which permits analysis of the size and shape of the 
average underlying solar cycle, as well as construction of the 
joint distribution of the size and timing of maximum daily
sunspot number.  We also show how the large variance in daily sunspot number
results in large variance in the monthly smoothed maximum sunspot number 
$\langle R \rangle_{\textrm{max}}$, and discuss the implications of this for 
the reliability of any forecast of the maximum of the cycle.  Section \ref{section:Forecasting} illustrates the
Bayesian framework from Section \ref{section:BayesianApproach} in
application to two historical solar cycles, namely cycle 19 (in Section
\ref{ssection:Cycle19}) and cycle 20 (in Section
\ref{ssection:Cycle20}).  Section \ref{section:Cycle24} applies the
Bayesian method to forecast the current cycle (cycle 24). Techniques from 
Section \ref{section:BayesianApproach} are used to
quantify the size of large/small sunspot numbers during cycle 24, and to
estimate the likely size and timing of the next maximum in both daily sunspot
number, and monthly smoothed sunspot number $\langle R \rangle_{\textrm{max}}$.

\section{A Bayesian Approach to Solar Cycle Forecasting}
\label{section:BayesianApproach}

\subsection{Fokker--Planck Model for Sunspot Number}
\label{ssection:FPModel}
The solar cycle variation in sunspot number comprises long--term 
secular variation, and large short--term statistical fluctuations.  
The long--term variation may be considered the underlying solar cycle, 
driven by the internal dynamo, and the short--term fluctuations 
attributed to complicated physical processes on the solar surface 
associated with sunspot formation, evolution and dispersion 
\cite{Parker1955}. 

The long--term solar cycle variation over a single cycle may be 
described by a cycle amplitude, cycle period, and cycle asymmetry 
\cite{HathawayEtAl1994}, which we represent with a `driver function' 
$\theta(t)$.  For illustrative purposes we consider first the simple 
choice for $\theta(t)$ of harmonic variation: 
\begin{equation} \label{eq:Sinedriver} \theta(t) = \alpha_0 +
\alpha_1 \sin \left( 2 \pi t / \alpha_2  + \alpha_3 \right),
\end{equation} 
where $\alpha_1$ is the cycle amplitude, $\alpha_2 \approx 11$ years is
the cycle period and $\alpha_3$ is the cycle phase.  With this choice
there is no cycle asymmetry.  

Short--term fluctuations in sunspot number on top of the driver function
$\theta(t)$ may be modelled using a probability distribution function
$f(s,t)=f(s,t|s_0,t_0)$, such that $f(s,t)ds$ is the probability that
the sunspot number is between $s$ and $s+ds$ at time $t$.  This approach
represents the sunspot number as a continuous random variable.
\ocite{NobleEtAl2011} modelled the time evolution of $f(s,t)$ using the
Fokker--Planck equation:
\begin{equation} \label{eq:FPE} 
\pd{f}{t} = \frac{1}{2}\pdd{}{s}\left[
\sigma^2(s,t)f(s,t)\right] - \pd{}{s} \left[ \mu(s,t) f(s,t)  \right],
\end{equation} where $\mu(s,t)$ describes deterministic changes in
sunspot number, and $\sigma^2(s,t)$ is a variance describing stochastic
variation.  Sunspot number is non--negative, so the appropriate boundary
condition at $s=0$ is a `zero probability flux' condition
\begin{equation} 
\label{eq:zeroflux}
\mu(s,t)f(s,t) -\left. \frac{1}{2}\pd{}{s}\left[
\sigma^2(s,t) f(s,t) \right] \right|_{s=0} = 0. 
\end{equation} 
An appropriate choice for $\mu(s,t)$, in terms of the driver function
$\theta(t)$ is \begin{equation} \label{equation:mr}\mu(s,t) = \kappa 
\left[ \theta(t) - s \right]. \end{equation} This choice causes the 
fluctuating sunspot
numbers to tend to return to the value $\theta(t)$ with a characteristic
timescale $1/\kappa$.  In this way the driver function $\theta(t)$ 
represents the secular or long--term sunspot number.  

The size of the observed squared deviations $r^2(t)=[s(t)-s(t-\Delta t)]^2$,
a proxy for daily sunspot number variance, tends to increase with sunspot
number \cite{NobleEtAl2011}.  Therefore a simple choice for the variance, 
which models this increase is
\begin{equation} \label{eq:Var} 
\sigma^2(s,t) = \beta_0 + \beta_1 s +
\beta_2 s^2, 
\end{equation} 
where $\beta_0$ describes variance in sunspot number at $s=0$, and
$\beta_1$ and $\beta_2$ describe the increase in variance with sunspot
number (we assume that $\beta_0, \beta_1$ and $\beta_2$ are all positive
or zero).  The model then has four parameters ($\kappa,\beta_0,\beta_1$
$\beta_2$), together with any parameters in the driver function
$\theta(t)$.  

The choices of Equations (\ref{eq:zeroflux}), (\ref{equation:mr}) and 
(\ref{eq:Var}) are 
discussed in detail in \ocite{NobleEtAl2011}.  The authors
showed that the model given by Equations (\ref{eq:Sinedriver}) to 
(\ref{eq:Var}) applied to historical monthly sunspot number data 
generates a probability distribution
function $f(s,t)$ which agrees both quantitatively and qualitatively
with observed sunspot statistics, even for the simple harmonic choice
for $\theta(t)$.  The model parameters were estimated from the
historical data using a maximum likelihood procedure explained in Section
\ref{ssection:Bayesian}.

A more realistic choice of driver function $\theta(t)$ for a single
solar cycle than Equation (\ref{eq:Sinedriver}) is provided by the
functional form \cite{HathawayEtAl1994}: 
\begin{equation}
\label{eq:Hath} \theta(t) = \frac{a \left(t-t_0\right)^3}{\exp\left[
-\left(t-t_0\right)^2/b^2 \right] - c}, 
\end{equation} 
where $t_0$ is the start of the cycle, and $a,b$ and $c$ represent the
cycle amplitude, period, and asymmetry respectively.  With this choice
of driver function there are seven parameters in the Fokker--Planck
model, which may be represented in a vector \begin{equation} \mb{\Omega}
= \left[a,b,c,\kappa,\beta_0,\beta_1,\beta_2 \right]. \end{equation} The
distribution of model sunspot numbers is written $f(s,t;\mb{\Omega})$ to
indicate the explicit dependence of the distribution on the model
parameters.  We assume that the cycle start date $t_0$ is known, 
but it could be treated as another parameter and estimated from the data.

If the parameters $\mb{\Omega}$ generating sunspot data are
known, the time evolution of the distribution of sunspot numbers
$f\left(s,t;\mb{\Omega}\right)$ is uniquely determined by the
Fokker--Planck Equation (\ref{eq:FPE}).  However, the parameters are
unknown.  To describe historical sunspot numbers the parameters may be
estimated from historical sunspot data, following \ocite{NobleEtAl2011}
[who used Equation (\ref{eq:Sinedriver}) as the choice of $\theta(t)$].
For forecasting, it is necessary to estimate values
$\hat{\mb{\Omega}}$ of the model to forecast future sunspot
numbers.  These procedures are explained in Section \ref{ssection:Bayesian}.   

\subsection{Bayesian Estimation of Model Parameters}
\label{ssection:Bayesian} Given an observed set 
$\mb{s} = \left\{s_0,s_1,\ldots,s_T \right\}$ of sunspot numbers at 
times $\left\{t_0,t_1,\ldots,t_T\right\}$, the maximum likelihood (ML)
estimate is the parameter set $\hat{\mb{\Omega}}$ which maximises
the likelihood function representing the probability of the data given 
the model:
\begin{equation}
\label{eq:Like}
\mathcal{L}\left(\mb{s}|\mb{\Omega}\right) = 
\prod_{i=1}^{i=T}f\left(s_i,t_i|s_{i-1},t_{i-1};\mb{\Omega}\right). 
\end{equation}
We assume that the distribution of $s_i$ at time $t_i$ depends
only on the previous observation $s_{i-1}$ at time $t_{i-1}$, which is
the Markov property \cite{KaratzasEtAl1988}.  ML estimates are optimal
in the sense that they are both efficient and consistent in large samples
\cite{DacunhaCastelleEtAl1986}.  However, ML estimates are limited in
that they only use information from the observed data, and ignore other 
information which may be available. 

With sunspot data we have additional information about the possible
size, shape, and length of a future solar cycle, which may be included
in the forecast using the Bayesian method ({\it e.g.} \opencite{Sivia2006}).  
The additional information may be in the form of a
dynamo--based forecast, or a precursor forecast, for a future cycle. 
Our confidence in the reliability of this information is represented
by a `prior distribution' $\mathcal{P}\left(\mb{\Omega}\right)$ for the
model parameters $\mb{\Omega}$ given the information.  For example, if a
precursor forecast for the variance parameter $\beta_0$ is
$\bar{\beta_0}$, and if the parameter $\beta_0$ is not correlated with
other model parameters, then an appropriate choice of a prior distribution 
for this parameter is
\begin{equation} \label{eq:UNormalPrior}
\mathcal{P}\left(\beta_0\right) = \frac{1}{\sqrt{2 \pi
\sigma^{2}_{\beta_0}}} \exp \left[-\frac{1}{2}\left( \frac{\beta_0 -
\bar{\beta}_0}{\sigma_{\beta_0}} \right)^2  \right], 
\end{equation}
where the variance $\sigma^{2}_{\beta_0}$ represents how confident 
we are that $\beta_0$ coincides with $\bar{\beta_0}$.  

Because we are dealing with multiple parameters which may be correlated,
it is necessary to include possible correlations in the prior.  For the 
choice of Gaussian--distributed priors it is appropriate to use the general
multinormal distribution 
\begin{equation} \label{eq:MNormalPrior}
\mathcal{P}\left( \mb{\Omega} \right) = \frac{1}{\left( 2 \pi
\right)^{k/2}|\Sigma|^{1/2}} \exp\left[-\frac{1}{2}\left( \mb{\Omega} -
\bar{\mb{\Omega}} \right) \Sigma^{-1} \left( \mb{\Omega} -
\bar{\mb{\Omega}} \right)'  \right], 
\end{equation} 
where $k$ is the number of parameters in $\mb{\Omega}$, and the matrix
$\Sigma$ is the variance--covariance matrix describing the uncertainties
in each parameter and the co--dependence of the parameters.  For the
parameters in the \ocite{HathawayEtAl1994} driver function (Equation 
(\ref{eq:Hath})), $\Sigma$ is a $7\times7$ matrix of the form
\begin{equation}
\Sigma =
\left( {\begin{array}{c c c c}
 \sigma^{2}_{a}  & \sigma^{}_{a,b} & \ldots & \sigma_{a,\beta_2}  \\
 \sigma^{}_{b,a} & \sigma^{2}_{b}  & \ldots & \sigma_{b,\beta_2}  \\
     \vdots      &    \vdots       & \ddots &  \vdots \\
 \sigma^{}_{\beta_2,a} & \sigma^{}_{\beta_2,b}  & \ldots & \sigma^{2}_{\beta_2}   
 \end{array} } \right),
\end{equation}
where $\sigma^{2}_{i}$ is the uncertainty in parameter $i$ and 
$\sigma_{i,j}$ is the covariance between parameters
$i$ and $j$, for $i,j = a,b,\ldots,\beta_2$.  As an example of the importance of 
correlations, it is well known that cycles which rise rapidly
tend to be large \cite{Waldmeier1935}, so that we expect parameters 
describing the period and amplitude to be negatively correlated.

Predictions incorporating prior information may then be made as follows. 
When a cycle begins, daily sunspot data $\mb{s} =
\left\{s_0,s_1,\ldots,s_T\right\}$ becomes available.  This can be
combined with the prior information by calculating the `posterior
distribution' $\mathcal{P}\left(\mb{\Omega}|\mb{s}\right)$, according to
Bayes' rule \cite{Sivia2006}:
\begin{equation}
\label{eq:Bayes}
\mathcal{P}\left( \mb{\Omega}|\mb{s} \right) =
\frac{\mathcal{P}\left(\mb{s}|\mb {\Omega}\right)\mathcal{P}\left(
\mb{\Omega} \right)}{\mathcal{P}\left(\mb{s} \right)}.
\end{equation}
The term $\mathcal{P}\left(\mb{s}|\mb{\Omega}\right)$ in Equation
(\ref{eq:Bayes}) is the likelihood function (\ref{eq:Like}), and the
denominator is a normalising constant.  The posterior distribution
combines information about $\mb{\Omega}$ contained in the data (the time
series approach), with relevant information external to the data ({\it e.g.}
from precursor and/or dynamo models, or any other source).  

We are unable to solve the Fokker--Planck equation (\ref{eq:FPE})
analytically, so we cannot evaluate the likelihood
$\mathcal{P}\left(\mb{s}|\mb{\Omega}\right)$ in closed form. An analytic
approximation appropriate for daily data is \cite{NobleEtAl2011}:
\begin{eqnarray}
\label{eq:SunspotSimDist}
f(s,\tau|s_0;\mb{\Omega}) & = & \frac{1}{\sqrt{2 \pi \sigma^2(s_0,t_0)
\tau}} \left[ \exp \left\{-\frac {\left[s-\left(s_0+\mu(s_0,t_0)\tau
\right)\right]^2}{2 \sigma^2(s_0,t_0) \tau }\right\} \right. \nonumber
\\ & & \quad \left. +\exp \left
\{-\frac{\left[s+\left(s_0+\mu(s_0,t_0)\tau \right)\right]^2}{2
\sigma^2(s_0,t_0)\tau} \right\} \right], 
\end{eqnarray}
where $\tau=t-t_0$.  Equation (\ref{eq:SunspotSimDist}) is the
conditional probability distribution function of the random variable 
$|s(t)|$, where $s(t)$ is a normal random variable with mean 
$s_0 + \mu(s_0,t_0)$ and variance $\sigma^2(s_0,t_0)\tau$, and $s_0$ 
is the sunspot number at time $t_0$.  Using this approximate solution, 
the likelihood function (\ref{eq:Like}) is 
\begin{eqnarray}
\label{eq:likelihood}
\mathcal{P}\left(\mb{s}|\mb{\Omega}\right) & = & \left( 2 \pi
\right)^{-\left(\frac{T}{2}\right)} \prod_{i=1}^{T} \left[ \exp
\left\{-\frac {\left[s_{i}-\left(s_{i-1}+\mu_{i-1} \tau
\right)\right]^2}{2 \sigma^{2}_{i-1}\tau }\right\} \right. \nonumber \\
& & \quad \left. +\exp \left \{-\frac{\left[s_{i}+\left(s_{i-1}+\mu_{i-1}
\tau \right)\right]^2}{2 \sigma^{2}_{i-1}\tau} \right\} \right],
\end{eqnarray}
where $\mu_i = \mu(s_i,t_i)$ and $\sigma_{i}^{2} = \sigma^2(s_{i},t_{i})$.

Given the posterior distribution $\mathcal{P}\left( \mb{\Omega}|\mb{s}
\right)$ a specific estimate for the parameters $\mb{\Omega}$ may be
calculated in a number of ways \cite{Jaynes2003}. In this
paper we use the the most probable, or `modal estimate' of $\mb{\Omega}$:
\begin{equation}
\label{eq:modal}
\hat{\mb{\Omega}} = \textrm{argmax} \hspace{+0.1cm} \mathcal{P}
\left( \mb{\Omega}|\mb{s} \right).
\end{equation}
\subsection{Construction of a Mean Solar Cycle Prior}
\label{ssection:mean}
Rather than using a specific precursor forecast, we consider using an 
historical average solar cycle, which we refer
to as a `mean cycle', as a prior. This means that before the start of a
cycle the most probable shape of the cycle ({\it i.e.} the parameters in the
driver function) and the variance of the cycle ({\it i.e.} the parameters in
$\sigma^2(s,t)$ of the coming cycle) are represented by an historical
average.  This choice may be interpreted as a `guess in total
ignorance'.       

To determine the parameters $\bar{\mb{\Omega}}$ for the mean cycle we
consider daily sunspot data for the previous 13 solar cycles over the
interval 1850 to 2010.  The \ocite{HathawayEtAl1994} driver function
given by Equation (\ref{eq:Hath}) is assumed to represent the shape of
each underlying cycle, and the variance of each cycle is modelled by
Equation (\ref{eq:Var}).  For each solar cycle maximum likelihood
estimates $\hat{\mb{\Omega}} = \left[\hat{a},\hat{b},
\hat{c},\hat{\kappa},\hat{\beta}_0,\hat{\beta}_1,
\hat{\beta}_2 \right]$ of the seven model parameters are calculated,
as shown in Table \ref{table1}. 
  
The average for each parameter over the previous 13 cycles is assumed to 
represent the mean cycle, and is used in our prior distribution.  
The sample means (denoted $\bar{\mb{\Omega}}$) are given in Table 
\ref{table2}.
The variance--covariance matrix $\Sigma$ is calculated using sample
covariances between the ML estimates for the seven parameters in Table
\ref{table1}.  For example the covariance between the
amplitude $a$ and period $b$ in $\Sigma$ is 
\begin{equation}
\sigma_{a,b} = \frac{1}{12}\sum_{i=1}^{13}\left( \hat{a}_i - \bar{a}
\right)( \hat{b}_i - \bar{b} ). 
\end{equation}
The (non-dimensional) correlation matrix is given in Table \ref{tablecorr}.
A number of important correlations exist.  In particular, the large 
correlation between $\kappa$ and $\beta_1$ (89\%) shows the strong 
relationship between the variance parameters and the rate at which
sunspot number returns to the level $\theta(t)$.  There are also 
significant correlations between the size of the cycle $a$, and the 
time to maximum $b$ and asymmetry $c$ (collectively describing the 
Waldmeier Effect \cite{Waldmeier1935}). 

\subsection{Fokker--Planck Modelling of the Mean Solar Cycle}
\label{ssection:Simulation}
In this section we investigate characteristics of the average solar
cycle using the mean cycle parameters $\bar{\mb{\Omega}}$ estimated from
daily sunspot data over the interval 1850 to 2010, given in Table
\ref{table2}.  The driver function Equation (\ref{eq:Hath}) with
parameter values from Table \ref{table2} describes the average
size and shape of the underlying solar cycle.  Short--term
deviations in sunspot number from this average are described by the
three variance parameters in Table \ref{table2}. 

\begin{figure}
 \centerline{\includegraphics[width=0.75\textwidth,clip=]{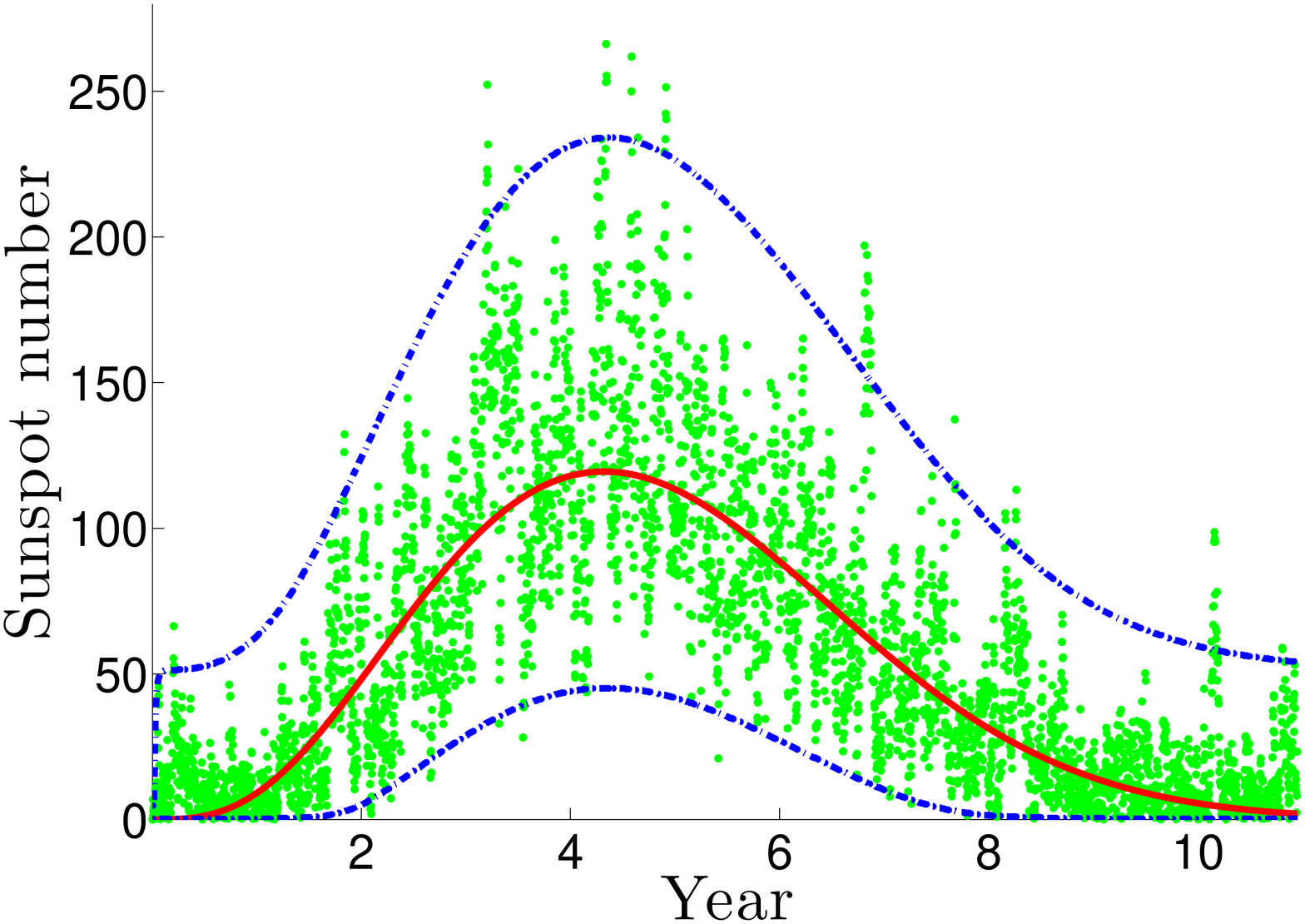}
      }
      \caption{Fokker--Planck modelling of the mean solar cycle, showing 
      the driver function (solid red curve), upper and lower 1\% quantiles 
      blue dot--dashed curve), and an example simulation of sunspot numbers 
      (green points), as described in Section \ref{ssection:Simulation}.  
      The maximum value attained 
      by the upper $1\%$ quantile is $s_{U}(t)=234$, which occurs 4.3 
      years after the start of the cycle.  In this simulation the maximum
      of the daily sunspot number is $s^* = 266$, which occurs approximately 
      4.4 years after the cycle begins.  }
\label{f1}
\end{figure}

Figure \ref{f1} illustrates the mean cycle ({\it i.e.} the
sunspot model with $\mb{\Omega}= \bar{\mb{\Omega}}$), and simulations of
daily sunspot number over the mean cycle using the Fokker--Planck model.  
The red curve (solid) in Figure \ref{f1} is the \ocite{HathawayEtAl1994}
driver function $\theta(t)$, given by Equation (\ref{eq:Hath}) with mean
cycle parameters $\bar{a},\bar{b}$ and $\bar{c}$.  The maximum value of
the driver function $\theta_{\textrm{max}}=119$ occurs 4.4 years after
the cycle start date.  To investigate the likely size of short--term
deviations about the driver $\theta(t)$, we numerically solve the
Fokker--Planck equation (\ref{eq:FPE}) for the initial initial condition
$f(s_0,t_0;\bar{\mb{\Omega}}) = \delta(s_0)$ with $s(t_0)=0$ at $t_0 =
0$.  The solution is obtained from $t=0$ to $t=11$ years.  Based on
the numerical solutions for $f(s,t|s_0,t_0;\bar{\mb{\Omega}})$ we
calculate the upper and lower 1\% quantiles, which are the curves
$s_U(t)$ and $s_L(t)$ defined by
\begin{equation}
\label{eq:CI}
\int_{0}^{s_L(t)}ds' f\left(s',t|s_0,t_0;\bar{\mb{\Omega}}\right) =
\int_{s_U(t)}^{\infty}ds' f\left(s',t|s_0,t_0;\bar{\mb{\Omega}}\right) =
0.01.
\end{equation}
These curves delineate boundaries of extreme sunspot numbers ({\it i.e.} the
probability that the sunspot number is larger or smaller than
$s_U(t)$ or $s_L(t)$ respectively, from a given initial condition is 1\%).  
The blue curves (dot/dashed) in Figure \ref{f1} show these upper and lower
1\% quantiles for the mean cycle.  The maximum value attained by the
upper $1\%$ quantile is $s_{U}(t) = 234$, which occurs 4.3 years after
the start of the cycle.  This means that on average solar maximum occurs
4.3 years after the start of the cycle, and that there is a 1\% chance
of observing a daily sunspot number larger than 234 at solar maximum for an
average cycle, given $s(t_0)=0$.  

Using the analytic approximation (Equation (\ref{eq:SunspotSimDist})) we
can simulate daily sunspot numbers over the mean cycle.  To do this we
repeatedly generate random variables $s(t)$ from the conditional probability
distribution (Equation (\ref{eq:SunspotSimDist})) with parameter values from 
Table \ref{table2}.  These numbers represent a sequence of possible daily
sunspot numbers (one Monte Carlo simulation).  The green points in
Figure \ref{f1} are an example of simulated daily sunspot
numbers. The maximum daily sunspot number $s^*$ for this particular
simulation is $s^*=266$, which occurs 4.4 years after the cycle begins. 
In this simulation 1.1\% and 1.8\% of the sunspot numbers are greater
than and less than the upper and lower 1\% quantiles respectively.   

\begin{figure}[ht!]
 \centerline{ \includegraphics[width=0.75\textwidth,clip=]{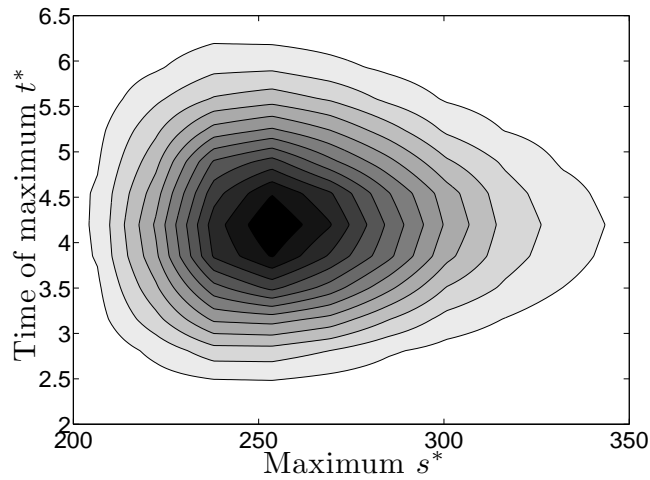} 
      }
      \caption{The joint distribution of daily maximum sunspot number $s^*$ 
      and time of maximum $t^*$ for the $5\times10^5$ Monte Carlo 
      simulations of the mean cycle described in Section 
      \ref{ssection:Simulation}.  The expected value of the maximum is 
      $271$, which occurs approximately $4.4$ years after the start of 
      the cycle.  The largest daily maximum value in any simulation 
      is 504, which suggests that the mean cycle has the potential 
      to generate extremely large sunspot numbers, although it is 
      very unlikely.} 
\label{f2}
\end{figure}

With the Fokker--Planck model we can investigate the likely size and
timing of daily sunspot maximum using repeated Monte--Carlo simulations.
We denote by $t^*$ the time of the occurrence of the maximum $s^*$ of the 
daily sunspot number, for one simulation.  Figure
\ref{f2} shows a Monte Carlo estimate of the joint
distribution $f(s^*,t^*|s_0,t_0;\bar{\mb{\Omega}})$ of the size and
timing of daily sunspot maximum based on $5\times10^5$ simulations. 
Each simulation is generated in the same way as the single instance
shown in Figure \ref{f1}.  The expected size of daily
sunspot maximum (the average over the simulations) is $\langle s^*
\rangle = 271$, which occurs approximately $\langle t^* \rangle = 4.4$
years after the start of the cycle.  This is comparable to the sample
average maximum sunspot number from the previous 13 cycles (see Section
\ref{ssection:mean}), which is $255$.   The largest value of the daily 
sunspot number of the $5\times10^5$ simulations is $s^*=504$, which 
suggests that mean cycle can generate extremely large 
sunspot numbers, although it is very unlikely (the mean cycle is 
expected to generate one such event every $5\times10^5$ cycles, 
or 5 million years).

\begin{figure}[ht!]
 \centerline{ \includegraphics[width=0.75\textwidth,clip=]{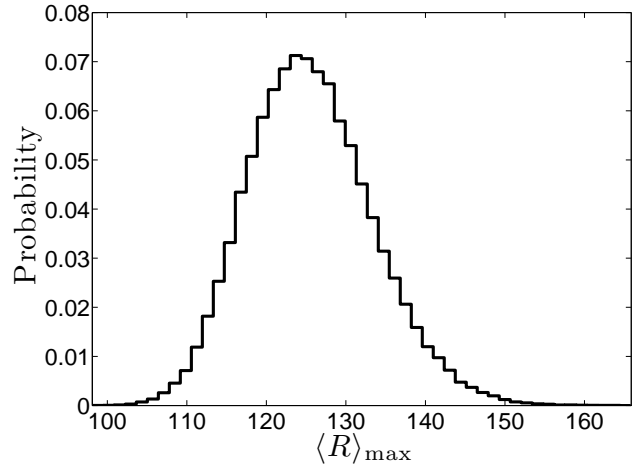} 
      }
      \caption{Distribution of the maximum of the 13 month smoothed
      sunspot number $\langle R \rangle_{\textrm{max}}$ calculated using 
      $5\times10^5$ simulations of daily sunspot number for the 'mean
      cycle' (see Section \ref{ssection:mean}).  The expected value of 
      the maximum is $\langle R \rangle_{\textrm{max}} = 125\pm8$.  
      The upper and lower 5\% quantiles are 113 and 138 
      respectively. } 
\label{fn}
\end{figure}

We can also investigate the monthly smoothed sunspot number 
$\langle R \rangle_{\textrm{max}}$, which is the main focus of much
of the existing literature \cite{Petrovay2010}.  During each simulation
of daily sunspot numbers discussed above, we calculate a 13--month
boxcar average sunspot number $\langle R \rangle$, and store the size of
the maximum $\langle R \rangle_{\textrm{max}}$.  Figure \ref{fn} plots
the distribution of $\langle R \rangle_{\textrm{max}}$ based on 
$5\times10^5$ simulations of daily sunspot number.  The expected value 
over the simulations is $\langle R \rangle_{\textrm{max}}=125 \pm 8$, 
and the lower and upper 5\% quantiles are are 113 and 138 respectively.  
For comparison the average smoothed maximum from the previous 13 cycles 
is 121.  

Figure \ref{fn} has important implications for any forecast of 
$\langle R \rangle_{\textrm{max}}$.  Even if the model parameters for 
a solar cycle are known, large daily variation in sunspot number 
causes large variation in the possible smoothed maximum value of the 
cycle.  This indicates that the reliability of any forecast of 
$\langle R \rangle_{\textrm{max}}$ is limited by the large daily 
fluctuations in sunspot number.

\section{Forecasting Historical Solar Cycles}
\label{section:Forecasting}
\begin{figure}[ht!]
\centerline{\includegraphics[width=0.75\textwidth,clip=]{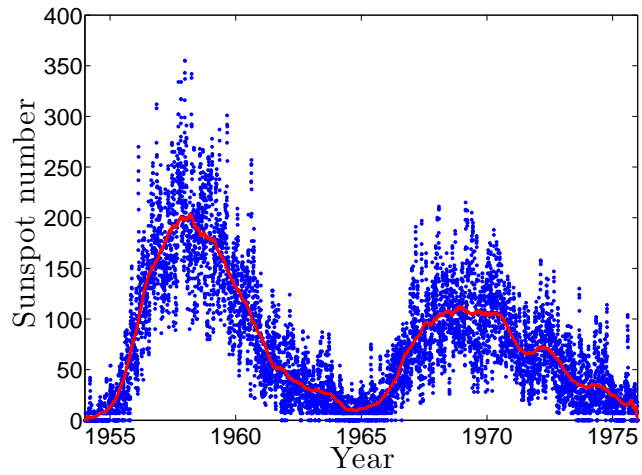} 
     }
	\caption{The daily sunspot number observations for cycles 19 and 20 
	(1954 to 1976) used for
	forecasting in Section \ref{section:Forecasting}.  The red curve is a
	smoothed sunspot number.  The maximum of the smoothed sunspot number 
	for cycle 19 is 203, and the maximum for cycle 20 is 113.}
\label{f3}
\end{figure}

In the following sections we consider application of the model to
forecasting two historical solar cycles: cycles 19 and 20. Cycle 19 is
chosen because it is the largest solar cycle since daily sunspot number
records began \cite{Kane2002}.  In contrast, Cycle 20 is very similar in
amplitude and shape to the mean cycle.  As such these cycles are very
different in size and shape, and useful as illustrations of the Bayesian
forecasting method described in Section \ref{ssection:Bayesian}.  The
two cycles are shown in Figure \ref{f3}.  The maximum of
the observed smoothed sunspot number for cycle 19 is 201, and the maximum for
cycle 20 is 110.   
    
In the forecasts in this section we use the mean cycle constructed in
Section \ref{ssection:mean} as a prior, and then we make updated
predictions using successively more historical data from the start of a
cycle, to demonstrate the Bayesian prediction method from Section
\ref{ssection:Bayesian}.  We use the analytic approximation (Equation
(\ref{eq:SunspotSimDist})) to the solution to the Fokker--Planck equation
to evaluate the likelihood function (\ref{eq:likelihood}), and
we use the mean cycle estimates discussed in Section \ref{ssection:mean}
with the multivariate normal prior distribution (Equation
(\ref{eq:MNormalPrior})).  Because $\mathcal{P}(\mb{s})$ in Bayes' rule
(\ref{eq:Bayes}) is required only as a normalising constant, we
calculate the posterior distribution $\mathcal{P}(\mb{\Omega}|\mb{s})
\propto \mathcal{P}(\mb{s}|\mb{\Omega}) \mathcal{P}(\mb{\Omega})$, and
evaluate the modal estimate $\hat{\mb{\Omega}}$ of Equation (\ref{eq:modal}) 
by numerical determination of the location of the maximum of the posterior
distribution.

\subsection{Solar Cycle 19}
\label{ssection:Cycle19}
Solar cycle 19, which occurred from 1954 to 1965, is the largest cycle on
record.  As such it provides a useful illustration of how
forecasts starting from a prior consisting of the mean cycle are
modified by observation of larger than expected sunspot numbers.  

First we consider applying the Bayesian estimation procedure to sunspot
data for the entire cycle.  If we take the mean cycle as the prior and
take all daily sunspot numbers from 1 January 1954 to 31 December 1964
as data $\mb{s}$, construction of the posterior
$\mathcal{P}(\mb{\Omega}|\mb{s}) \propto \mathcal{P}(\mb{s}|\mb{\Omega})
\mathcal{P}(\mb{\Omega})$ and estimation of parameters gives the modal
estimates $\hat{\mb{\Omega}}$ in Table \ref{table3}.  The difference 
between the ML estimates in Table \ref{table1} and the Bayesian estimates
in Table \ref{table3} is the influence of the prior distribution in the
calculation of the posterior.

\begin{figure}[ht!]
  \centerline{ \includegraphics[width=0.75\textwidth,clip=]{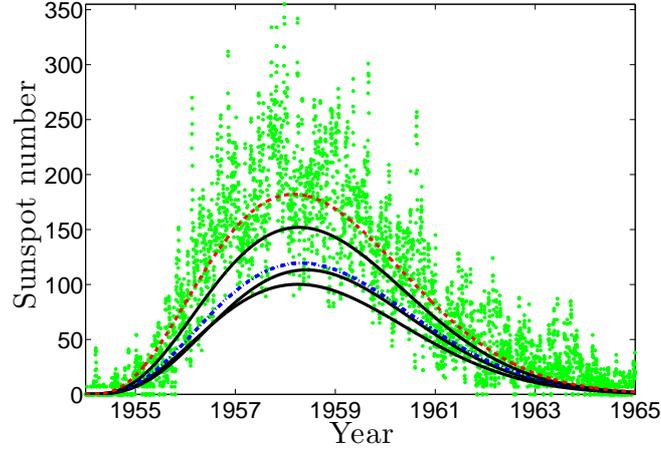} 
  }
  \caption{Prediction of cycle 19 using successively more data from the 
  start of the cycle.  Daily sunspot numbers for the cycle are shown by the 
  green points.  The driver function for the mean cycle prior which uses no 
  sunspot data is in blue (dot--dashed), and the Bayesian estimate using 
  all sunspot data for 1954 to 1964 is the red dashed curve.  The model 
  parameters for the red curve are given in Table  \ref{table3}.  The solid 
  black curve with the smallest maximum value is the forecast using ten days of 
  data, the next smallest uses one year of data, and the largest solid black 
  curve uses two years of data. }
\label{f4}
\end{figure}

We can repeat the calculation using only part of the data from the start
of cycle 19 as input in $\mb{s}$. By constructing the posterior repeatedly, 
using successively more data from the start of the cycle, we mimic the process
of forecasting and updating the forecast.  Figure \ref{f4} illustrates the 
process of successive forecasting for this cycle.  The green points are the 
observed daily sunspot numbers for cycle 19.  The driver function for the mean 
cycle (the prior for the forecasts) is shown in blue (dot/dash).  The three 
black curves (solid) are the driver functions given by Equation (\ref{eq:Hath}) 
calculated using Bayesian model parameters estimated with different amounts of 
daily sunspot data.  The black curve with the smallest maximum value is obtained 
using ten days of data from the start of the cycle, the next smallest uses one
year of data from the start of the cycle, and the black curve with the
largest maximum value uses two years of data.  The driver function
corresponding to the final Bayesian estimate using all data for the
cycle (which has a maximum of $\theta_{\textrm{max}}=182$) is shown by the 
red dashed curve.  This figure shows how, as parameter estimates are updated with
additional daily sunspot data, the size, period and asymmetry of the
forecast of the underlying solar cycle changes. Initial estimates of the
size of the sunspot maximum are lower than that of the mean cycle, but
the large sunspot numbers observed from about 1956 onwards cause the
estimates of the cycle maximum to increase.    

\begin{figure}[ht!]
\centerline{\includegraphics[width=0.75\textwidth,clip=]{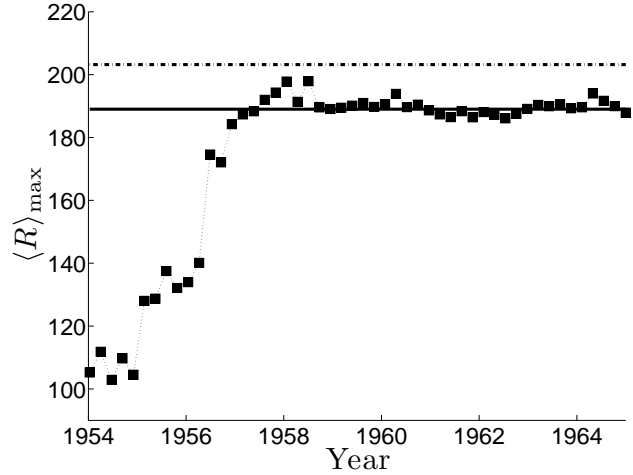} 
     }
	\caption{Prediction of cycle 19 using successively more data from 
	the start of the cycle.  The value of the maximum $\langle R 
	\rangle_{\textrm{max}} = 203$ for the observed cycle 19 data is 
	shown by the dot--dashed line.  The expected value of $\langle R 
	\rangle_{\textrm{max}}$ calculated by average over 10$^5$ cycles
	is $\langle R \rangle_{\textrm{max}} = 189$, and is shown by the 
	solid line.  The forecasts of $\langle R \rangle_{\textrm{max}}$ 
	for the Bayesian modal estimate using daily sunspot up to the indicated 
	time, and calculated by averaging over 10$^3$ simulations are shown 
	by the squares.  From 1955 to 1957 the forecast of 
	$\langle R \rangle_{\textrm{max}}$ rises rapidly and then is 
	approximately constant.  The final value, matching the parameters 
	in Table \ref{table3}, is $\langle R \rangle_{\textrm{max}}=182$.}
\label{f5}
\end{figure}

Figure \ref{f5} shows the estimate of maximum smoothed sunspot number
$\langle R \rangle_{\textrm{max}}$ as a function of the time of the 
latest data used for the prediction.  These estimates are calculated by
averaging over $10^5$ simulations (blue squares), as discussed in 
Section \ref{ssection:Simulation}.  The first estimate of 
$\langle R \rangle_{\textrm{max}}$ is calculated using 10 days of data 
which consisted of 10 consecutive spotless days.  The solid black curve is 
the expected value of $\langle R \rangle_{\textrm{max}}$ calculated using all 
daily data for cycle 19.   Early Bayesian estimates 
({\it i.e.} using data from 1954 to 1955) of
$\langle R \rangle_{\textrm{max}}$ are small because the data is dominated by a
large number of days early in the cycle with zero sunspot number. 
From 1955 onwards the sunspot numbers increase more rapidly than expected
for the mean cycle.  As a result the Bayesian result for
$\langle R \rangle_{\textrm{max}}$ rises rapidly until around 1958, and after
that the estimate of $\langle R \rangle_{\textrm{max}}$ is approximately 
constant, fluctuating between 180 and 195.  The final estimate ({\it i.e.} the estimate
using all daily sunspot data for cycle 19) is 
$\langle R \rangle_{\textrm{max}}=189 \pm 11$, corresponding to the parameters in
Table \ref{table3}.  The observed value of 
$\langle R \rangle_{\textrm{max}}=201$ is shown by the black dashed line, and is 
roughly one standard deviation higher than the expected value, given the data.  
This difference illustrates the large variability in the cycle maximum possible 
due to the daily sunspot number fluctuations (see Section \ref{ssection:Simulation}). 

\subsection{Solar Cycle 20}
\label{ssection:Cycle20}
Solar cycle 20, which occurred from 1965 to 1976, is substantially
different in character to cycle 19, discussed in Section
\ref{ssection:Cycle19}.  The shape of this cycle is more typical, similar 
to the mean cycle.

Following the approach in Section \ref{ssection:Cycle19}, we first
consider Bayesian estimation applied to daily sunspot data for the
entire cycle, using the mean cycle as a prior.  The data span 1 January
1965 to 31 December 1976.  Table \ref{table4} lists
the model estimates $\hat{\mb{\Omega}}$  for the Fokker--Planck
model parameters obtained using the Bayesian procedure from Section
\ref{ssection:Bayesian} applied to the daily sunspot data for the 
whole cycle.  The difference between
the Bayesian and ML estimates is again due to the influence of the prior
in the calculation of the posterior distribution.  

Figure \ref{f6} illustrates predictions for cycle 20
following Figure \ref{f4}.  The green points are the observed
daily sunspot numbers for cycle 20.  The three black (solid) curves show 
the driver function (Equation (\ref{eq:Hath})) calculated using successively
more sunspot data.  The black curve with the largest maximum is
estimated using ten days of data from the start of the cycle, 
the next largest uses one year of data, and the third black curve uses 
two years of data.  The driver
function corresponding to the final Bayesian estimate (which has a
maximum $\theta_{\textrm{max}}=124$) is shown by the red dashed curve.  
Initial estimates of the cycle amplitude are larger than that of the mean 
cycle, as indicated in Figure \ref{f6}.  The observation of many
days with small sunspot numbers ({\it i.e.} $s_i<50$) up to three
years into the cycle causes these large initial estimates of cycle
amplitude to be reduced.  The timing of the maximum of the cycle is
correspondingly adjusted from late 1969 to late 1968.    

\begin{figure}[ht!]
\centerline{\includegraphics[width=0.75\textwidth,clip=]{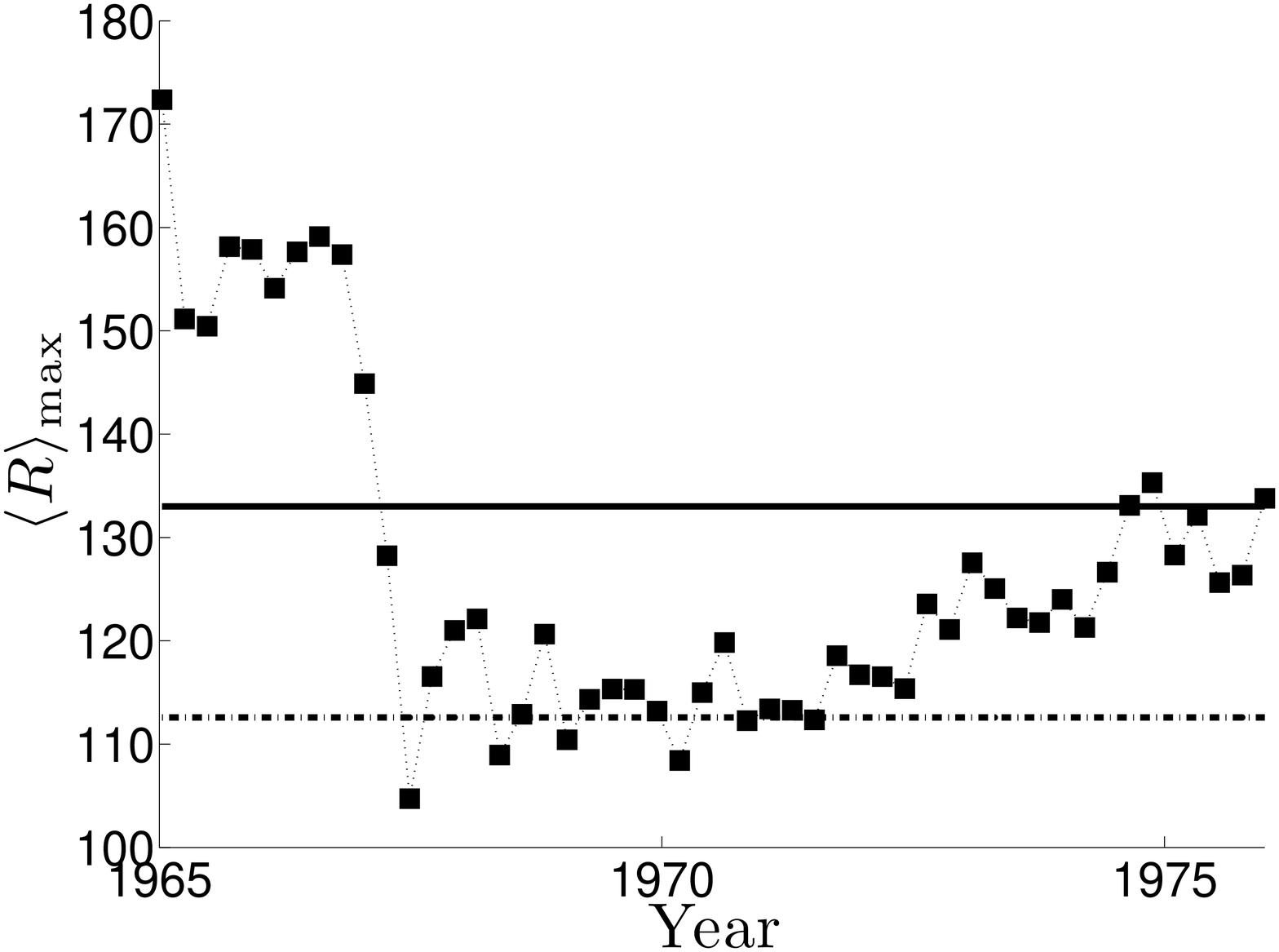} 
	}
	\caption{Prediction of cycle 20 using successively more data from 
	the start of the cycle.  The value of the maximum $\langle R 
	\rangle_{\textrm{max}}=113$ for the observed cycle 20 data is shown
	by the dot--dashed line.  The expected value of $\langle R 
	\rangle_{\textrm{max}}$ calculated by averaging over $10^5$ simulations
	is $\langle R \rangle_{\textrm{max}}=133$, shown by the solid line.  
	The forecasts of $\langle R \rangle_{\textrm{max}}$ for the Bayesian modal 
	estimate using daily sunspot data up to the indicated time, and calculated by 
	averaging over $10^3$ simulations are shown by the squares.  The 
	significant number of large sunspot numbers at the start of the cycle
	cause early estimates of $\langle R \rangle_{\textrm{max}}$ to be 
	larger than expected.  However, the lack of large sunspot numbers during
	solar maximum cause estimates of $\langle R \rangle_{\textrm{max}}$
	to be reduced.  The large variation in daily sunspot numbercauses estimates of 
	$\langle R \rangle_{\textrm{max}}$ to slowly rise to the final value 
	($\langle R \rangle_{\textrm{max}}=133$), matching the parameters in Table 
	\ref{table4}.}
	\label{f7}
\end{figure}

Figure \ref{f7} shows the estimate of maximum smoothed sunspot number  
$\langle R \rangle_{\textrm{max}}$ as a function of the time of the last
data used for the prediction (blue squares).  These estimates are calculated 
by averaging over $10^5$ simulations, as discussed in Section
\ref{ssection:Simulation}.  The solid black curve is the expected value of 
$\langle R \rangle_{\textrm{max}}$ calculated using all daily data for 
cycle 20.  In this case there are a significant number of days with
relatively large sunspot number at the start of the cycle (1965--1966),
which cause the initial estimates of $\langle R \rangle_{\textrm{max}}$ to 
be larger than that of the mean cycle. However, from mid 1966 onwards there
are many days with small sunspot numbers ({\it i.e.} $s_i<50$), and few days with 
large sunspot number ({\it i.e.} $s_i > \theta(t)$).  This causes the forecast to 
be reduced.  The declining phase of
cycle 20 (1969--1972) features a significant number of days with large
sunspot number, which causes the forecast to increase again.  The
final estimate of $\langle R \rangle_{\textrm{max}}$ using all daily 
sunspot data for cycle 20 is 
$\langle R \rangle_{\textrm{max}}=133 \pm 11$, corresponding to the 
parameters in Table \ref{table4}).  The observed value of 
$\langle R \rangle_{\textrm{max}}$ is 113, which is roughly two standard
deviations less than expected, given the data, again illustrating the
possible variability in the maximum value.     

\section{Forecasting the Current Solar Cycle (Cycle 24)}
\label{section:Cycle24}
In this section the Bayesian forecasting procedure is applied to forecasting
the current solar cycle, cycle 24.  There is considerable interest in
forecasts for the new cycle given its late start and slow early onset
\cite{RussellEtAl2010}. In particular the years 2008 and 2009 featured
long sequences of days in which the Sun was `spotless' \cite{TokumaruEtAl2009}, 
and various features of the new cycle have prompted speculation that future 
activity will be substantially reduced ({\it e.g.} 
\opencite{LivingstonEtAl2009})

Following Sections \ref{ssection:Bayesian}, \ref{ssection:mean}, and
\ref{ssection:Simulation}, the mean cycle is used as a prior.  The start
of cycle 24 is taken to be 1 January 2009.  With these assumptions the 
Bayesian estimates of the Fokker--Planck model parameters using all available 
daily sunspot data 1 January to 31 March 2011 are given in Table 
\ref{table5}.  The maximum value
$\theta_{\textrm{max}}$ of the driver function corresponding to the
parameters in Table \ref{table5} is 61, which is
approximately half the value for the mean cycle.  The available data
suggests that cycle 24 will be significantly smaller than average.   

Figure \ref{f8} illustrates the forecasts for cycle 24 following 
Figures \ref{f4} and \ref{f6}.  The daily sunspot numbers
for cycle 24 for the interval January 2009 to March 2011 are shown by
the green points, and the driver function for the mean cycle is shown by the
blue dot-dashed curve.  The solid black curve with the largest maximum value 
is the driver function
using the Bayesian estimate of the Fokker--Planck model based on the
first year of sunspot data (January 2009 -- January 2010), and the
second solid black curve uses the first two years of data.  Combining the mean
cycle prior with the first year of data gives estimates of the driver
function very similar to the driver function of the mean cycle.  However, due
to the large number of days during the latter part of 2010 with small
sunspot numbers, the driver function using the first two years of data
has a much smaller maximum $\theta_{\textrm{max}}$ than that of the mean
cycle.  The red dashed curve is the forecast using all data, which has a
maximum $\theta_{\textrm{max}}=61$.

\begin{figure}[ht!]
\centerline{\includegraphics[width=0.75\textwidth,clip=]{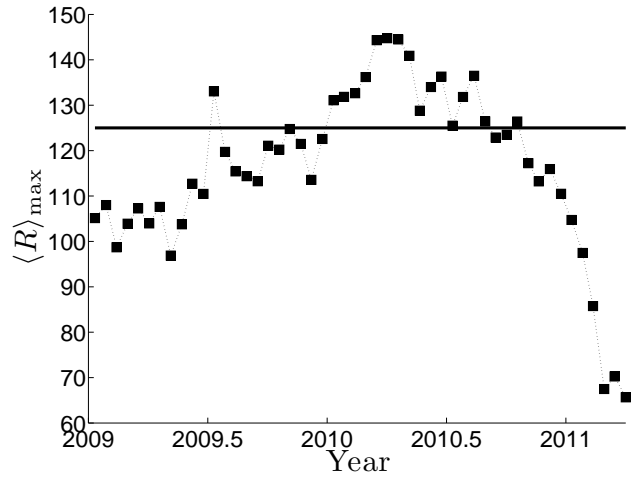} 
	}
	\caption{Prediction of cycle 24 using successively more data from 
	the start of the cycle.  The expected maximum $\langle R 
	\rangle_{\textrm{max}}=125$ for the mean cycle is shown by the solid
	line.  The forecasts of $\langle R \rangle_{\textrm{max}}$ for the
	Bayesian modal estimate using daily sunspot data up to the indicated
	time, and calculated by averaging over $10^3$ simulations are shown by
	the squares.  The lack of large sunspot numbers in late 2010 causes 
	a dramatic reduction in the expected size of 
	$\langle R \rangle_{\textrm{max}}$. The final value, matching the parameter 
	estimates in Table \ref{table5}, is $\langle R \rangle_{\textrm{max}}=66$. }
	\label{f9}
\end{figure}

Figure \ref{f9} shows the expected value of $\langle R \rangle_{\textrm{max}}$ 
as successively more data are incorporated into the Bayesian
prediction method starting from 1 January 2011, following Figures 
\ref{f5} and \ref{f7}. These (blue squares) estimates are calculated by 
averaging over $10^5$ simulations, as discussed in Section 
\ref{ssection:Simulation}.  The solid black curve is the expected maximum of 
$\langle R \rangle_{\textrm{max}}$ for the mean cycle.  The early forecasts of 
$\langle R \rangle_{\textrm{max}}$  are lower than that of the mean cycle because 
of the significant number of days during 2009 with zero sunspot number.  The 
forecasts of $\langle R \rangle_{\textrm{max}}$ steadily increase from early 
2009 until mid--2010, but sunspot activity defied expectation and did not significantly 
increase during the latter part of 2010, and this causes a  
dramatic reduction in the forecast for $\langle R \rangle_{\textrm{max}}$ during 
late 2010 and early 2011.  The final estimate using all available data 
(and matching the parameters in Table \ref{table5}) 
is 2009 is $\langle R \rangle_{\textrm{max}} = 66 \pm 5$.  This suggests that
cycle 24 will similar in size to cycle 14, and thus larger only than cycles
5 and 6.  This prediction is close to the smaller estimates in the literature. 
For example, \ocite{AguirreEtAl2008} predicted 
a smoothed sunspot maximum of $65 \pm 16$, \ocite{CameronEtAl2007} predicted 
a smoothed maximum of $69 \pm 15$, and \ocite{Kakad2011} a smoothed maximum 
of $74 \pm 10$.   

\begin{figure}[ht!]
	\centerline{\includegraphics[width=0.75\textwidth,clip=]{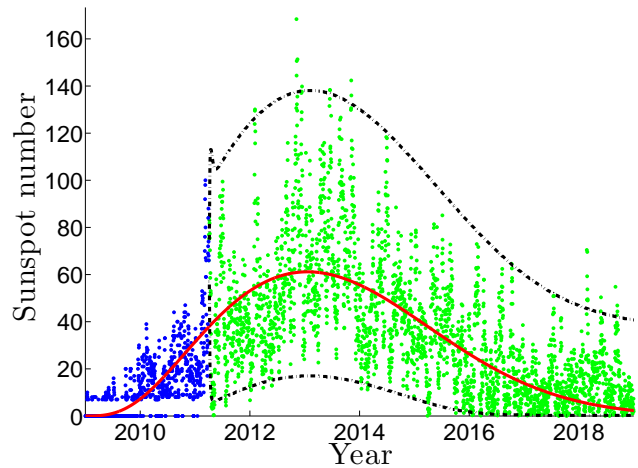} 
	}
	\caption{Prediction of cycle 24: illustration of the forecast 
	distribution of daily sunspot numbers for the remainder of cycle 
	24.  The model parameters used in the forecast are the Bayesian 
	estimates given in Table 
	\ref{table5}.  The solid red curve is the forecast 
	of the driver function, and the two dot--dashed black curves are the upper and 
	lower 1\% quantiles for the sunspot number distribution.  The blue 
	points are the daily sunspot numbers observed for January 2009 to 
	March 2011 used for the prediction.  The green points are a simulation 
	of future sunspot numbers using the parameters in Table 
	\ref{table5}.  The upper quantile attains a maximum value of 138 during the 
	period January--March 2013, identifying this as the most likely time for 
	a maximum in the daily sunspot numbers.}
\label{f10}
\end{figure}

Figure \ref{f10} provides a third representation of
the forecasts for cycle 24 based on the daily sunspot data for January 
2009 to March 2011 (the observed data are shown in blue).  The
solid red curve is the driver function forecast based on all observed data, 
matching the Bayesian model estimates in Table \ref{table5}.  From April 2011
to January 2019 the solid red curve provides a basis for prediction of the 
upcoming sunspot numbers.  The two dot--dashed black curves are the upper and 
lower 1\% quantiles for the sunspot number distribution for the forecast, defined 
by Equation (\ref{eq:CI}).  These quantiles show the probability of
excursions to large and small daily sunspot numbers.  The maximum value
attained by the upper 1\% quantile is 138 during the period
January--March 2013, which may be taken as the most likely time $t^*$
of daily sunspot maximum $s^*$.  The green points are simulated daily
sunspot numbers for the remainder of cycle 24 using the Bayesian model 
estimates in Table \ref{table5}, with initial condition $s=66$ on
March 31 2011 (the sunspot number observed on that day).  In this
particular simulation the maximum daily sunspot number is $s^*=168$ which 
occurs during October 2012.  For the simulation a total of 0.9\% and 1.1\% of
simulated sunspot numbers fall above and below the upper and lower 1\%
quantiles respectively.

\begin{figure}[ht!]
\centerline{\includegraphics[height=2.5in]{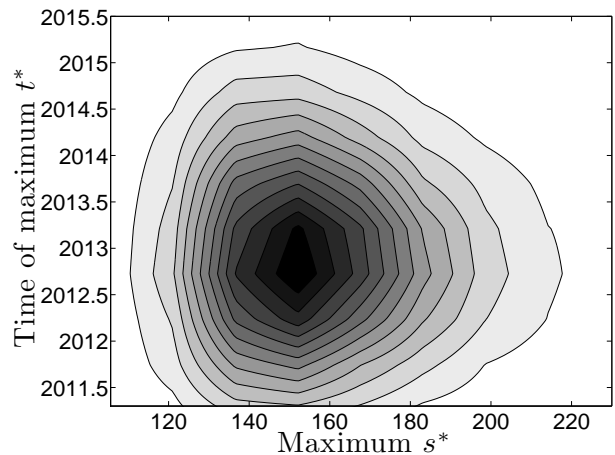} 
	}
	\caption{The joint distribution of sunspot maximum $s^*$ and 
	time of maximum $t^*$ for a sequence of $5\times10^5$ Monte 
	Carlo simulations of solar cycle 24, as described in Section 
	\ref{section:Cycle24}.  The simulation uses the Bayesian 
	model parameters in Table \ref{table5}, 
	and the initial condition $s_0=66$.  The expected size of the 
	daily sunspot maximum is $166\pm24$, which is most likely to occur 
	around March 2013.}
\label{f11}
\end{figure}

Figure \ref{f11} shows the joint distribution of the time
$t^*$ and size $s^*$ of daily sunspot maximum for cycle 24, generated
using $5\times10^5$ simulations of daily sunspot number based on the
Bayesian estimates in Table \ref{table5}. 
Averaging over the simulations we calculate the expected size of the
maximum daily sunspot number to be $\langle s^* \rangle = 166\pm24$, and 
this is expected to occur at a time $\langle t^* \rangle$ during
March 2013.  The sample average daily sunspot number maximum over the
previous 13 cycles is $\bar{s^*}=255$, so on this basis cycle 24 is
expected to be significantly smaller than average.  The model probability 
that daily sunspot number maximum for cycle 24 is larger than the average
maximum daily sunspot number $\bar{s^*}=255$ is $\mathcal{P}\left( s^* >
255 \right) = 0.4\%$.  Hence it is unlikely that individual days with
very large sunspot numbers will be observed during cycle 24.  

\section{Discussion}
\label{section:Discussion}
This paper introduces new techniques for estimating, analysing, and
forecasting solar cycles, in particular daily and smoothed sunspot 
numbers for a cycle, and their statistical properties.  In particular,
we have shown that even with perfect knowledge of the details of a solar
cycle, the observed sunspot maximum (either daily or smoothed) could
achieve a broad range of values due to the large fluctuations in the
daily sunspot number.  This is important for all prediction
done a priori, and indicates the true reliability of any forecast of the 
maximum of a cycle made before the fact. 

The main result of this paper is a new Bayesian prediction method for 
daily sunspot number (Section \ref{ssection:Bayesian}).  This method is
illustrated in application to two dissimilar historical cycles (Section 
\ref{section:Forecasting}) , and then is applied to the upcoming solar 
cycle (Section \ref{section:Cycle24}).  The method uses as a prior a 
mean cycle based on the observed solar cycles for 1850--2010 (Section 
\ref{ssection:mean}).  Our investigation of this provides a 
characterisation of solar cycle variability which should also be useful 
to other workers.        

We model the sunspot number as a continuous--time stochastic process,
with a probability distribution function described by a Fokker--Planck
equation \cite{NobleEtAl2011}. The Bayesian approach to forecasting uses
the Fokker--Planck model to include information about solar cycles
contained in sunspot data observed up to a given time, and combines these
with external information (in principle that provided by precursor or
dynamo--based forecasts).  The external information is included by
specifying an appropriate prior distribution. In this paper we take an
historical average solar cycle ( a `mean cycle') as a prior, which can
be interpreted as a `best guess in total ignorance'.  However, the
methodology can accommodate any choice of prior.  The Bayesian
estimation method, combined with the Fokker--Planck equation approach,
allows forecasts of the size and shape of the underlying solar cycle, as
well as assigning probabilities to the observation of large deviations
in sunspot number via calculation of upper and lower quantiles for
future sunspot numbers.  

In addition, the Fokker--Planck model permits
daily sunspot numbers to be simulated over a solar cycle, allowing Monte
Carlo construction of the joint distribution of the size and timing of
the maximum in daily sunspot number, as well as the distribution of the
size of smoothed sunspot maximum $\langle R \rangle_{\textrm{max}}$.  
In particular, the distribution of daily sunspot maximum determines 
the possible size and timing of extreme sunspot numbers during a cycle, 
which define likely times for the occurrence of intense solar activity. 
Large flares and coronal mass ejections occuring at these times are
drivers of our local space weather \cite{2008sswe.rept.....C}, and  
forecasting of extreme events space weather is an important 
task \cite{Petrovay2010}.     

The application of the new method to the current solar cycle, cycle 24,
provides insight into what we might expect over the next
few years.  Taking the mean solar cycle as  prior and using data for 1 
January 2009 to 31 March 2011, the model forecast for the maximum of the
smoothed sunspot number is $\langle R \rangle_{\textrm{max}}=66\pm5$, 
which is a very low value.  The forecast maximum daily sunspot number is 
$166\pm24$, expected to occur during March 2013, and this is also very low.  
These predictions are consistent with other predictions in the literature 
in suggesting a much smaller than average cycle.  The lack of a rapid rise 
in sunspot number during 2010, in particular, is shown by our modelling to 
imply a very small upcoming solar cycle.

\begin{acks}
We thank Don Melrose for his comments on the manuscript, and an anonymous 
reviewer whose detailed comments helped improve the paper.  P.~N. gratefully 
acknowledges a University of Sydney Postgraduate Scholarship.
\end{acks}

\begin{figure}[ht!]
\centerline{\includegraphics[width=0.75\textwidth,clip=]{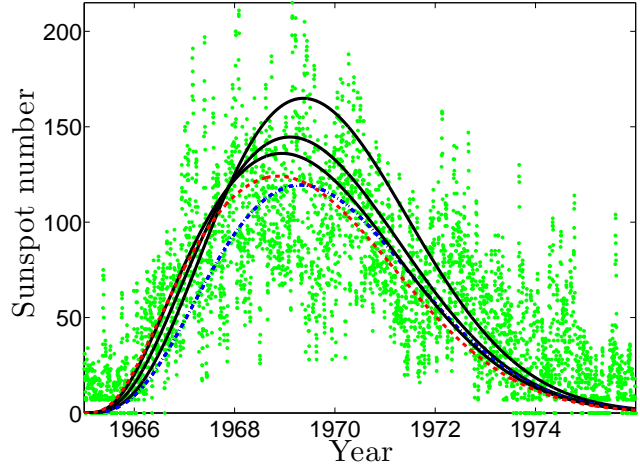} 
      }
	  \caption{Prediction of cycle 20 using successively more data from 
	  the start of the cycle.  Daily sunspot numbers for the cycle are 
	  shown in green.  The driver function for the mean cycle prior
	  which uses no sunspot data is the blue dot--dashed curve, and the 
	  Bayesian estimate using all sunspot data for 1965 to 1975 is 
	  shown by the red dashed curve.  The model parameters for the red dashed
	  curve are given in Table \ref{table4}.  The black solid curve with 
	  the largest amplitude is estimated using ten days of data, the next 
	  largest uses one year of data, and the largest solid black curve uses 
	  two years of data respectively. }
\label{f6}
\end{figure}

\begin{figure}[ht!]
	\centerline{\includegraphics[height=2.5in]{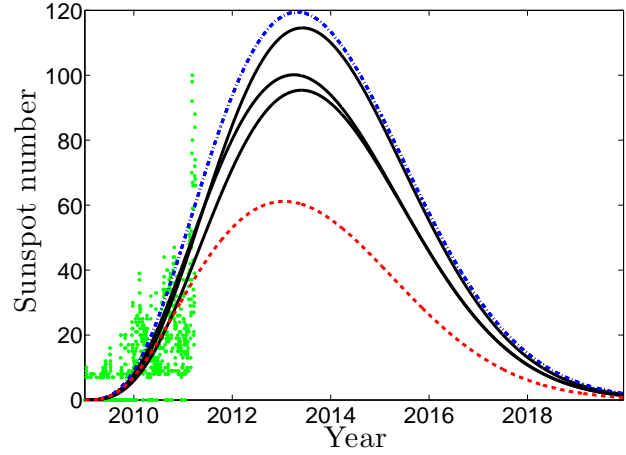} 
	}
	\caption{Prediction of cycle 24 using successively more data from 
	the start of the cycle.  Sunspot data for the cycle are shown by the 
	green points.  The driver for the mean cycle prior which uses no sunspot
	data is shown by the blue dot--dashed curve, and the driver function for 
	the Bayesian estimates using all available data for the cycle (January 
	2009 to March 2011) is shown by the red dashed curve.  The solid black 
	curve with the largest maximum 	value is the forecast for the driver 
	function using one year of daily sunspot data from the start of the cycle, 
	and the second solid black curve uses two years of data.}
	\label{f8}
\end{figure}

\begin{table}[ht!]
   \centering
   \begin{tabular}{ c c c c c c c c} 
   Cycle & $\hat{a}$ & $\hat{b}$ & $\hat{c}$ 
         & $\hat{\kappa}$ & $\hat{\beta}_0$ & $\hat{\beta}_1$ 
         & $\hat{\beta}_2$ \\
         & {\scriptsize $[\mbox{day}^{-3}]$}   & {\scriptsize $[\mbox{day}]$} 
         & & {\scriptsize$[\mbox{day}^{-1}]$}  & {\scriptsize$[\mbox{day}^{-1}]$} 
         & {\scriptsize $[\mbox{day}^{-1}]$} & {\scriptsize $[\mbox{day}^{-1}]$} \\
   \hline
23 & 7.8234$\times 10^{-8}$ & 1514.8 & 0.22174 & 0.08555 & 17.689 & 1.6569 & $2.1862 \times 10^{-5}$ \\
22 & 14.112$\times 10^{-8}$ & 1368.2 & 0.33153 & 0.07305 & 22.361 & 1.6302 & $1.1872 \times 10^{-3}$ \\
21 & 12.497$\times 10^{-8}$ & 1414.0 & 0.48998 & 0.07325 & 22.069 & 1.4652 & $3.0413 \times 10^{-3}$ \\
20 & 7.2861$\times 10^{-8}$ & 1450.4 & 0.90401 & 0.06151 & 44.288 & 1.1235 & $1.2022 \times 10^{-3}$ \\
19 & 14.048$\times 10^{-8}$ & 1391.4 & 0.66332 & 0.07994 & 18.636 & 1.8863 & $7.3281 \times 10^{-7}$ \\
18 & 10.592$\times 10^{-8}$ & 1411.5 & 0.65807 & 0.09238 & 22.541 & 2.3351 & $1.6617 \times 10^{-4}$ \\
17 & 7.2515$\times 10^{-8}$ & 1440.4 & 0.80478 & 0.11587 & 30.236 & 1.7644 & $7.4863 \times 10^{-3}$ \\ 
16 & 5.1614$\times 10^{-8}$ & 1457.8 & 0.43823 & 0.13072 & 14.619 & 2.8419 & $4.0631 \times 10^{-3}$ \\
15 & 7.8341$\times 10^{-8}$ & 1285.2 & 0.83355 & 0.10493 & 23.245 & 3.4815 & $2.8562 \times 10^{-3}$ \\
14 & 3.5130$\times 10^{-8}$ & 1576.8 & 0.42673 & 0.12617 & 10.136 & 3.5719 & $8.5559 \times 10^{-4}$ \\
13 & 8.5060$\times 10^{-8}$ & 1256.3 & 0.81241 & 0.13579 & 24.716 & 3.3131 & $2.4606 \times 10^{-4}$ \\
12 & 3.7627$\times 10^{-8}$ & 1526.9 & 0.55401 & 0.12793 & 14.543 & 3.2136 & $2.1382 \times 10^{-4}$ \\
11 & 9.7135$\times 10^{-8}$ & 1356.6 & 0.73723 & 0.17509 & 40.229 & 4.5763 & $9.1951 \times 10^{-4}$ \\
   \end{tabular}
   \caption{Maximum likelihood estimates of the parameters $\mb{\Omega} = 
   \left[a,b,c,\kappa, \beta_0,\beta_1,\beta_2 \right]$ for the previous 
   13 solar cycles over the interval 1850 to 2010.}
   \label{table1}
\end{table}

\begin{table}[ht!]
   \centering
   \begin{tabular}{ c c c c c c c} 
   $\bar{a} $ & $\bar{b}$ & $\bar{c}$ & $\bar{\kappa}$ & $\bar{\beta}_0$ & 
   $\bar{\beta}_1$ & $\bar{\beta}_2$ \\
      {\scriptsize $[\mbox{day}^{-3}]$}   & {\scriptsize $[\mbox{day}]$} 
   & & {\scriptsize$[\mbox{day}^{-1}]$}  & {\scriptsize$[\mbox{day}^{-1}]$} 
     & {\scriptsize $[\mbox{day}^{-1}]$} & {\scriptsize $[\mbox{day}^{-1}]$} \\
   \hline
   8.6233$\times 10^{-8}$ & 1419.3 & 0.60582 & 0.10633 & 23.486 & 2.5277 & 1.7123 $\times 10^{-3}$
   \end{tabular}
   \caption{Sample means $\bar{\mb{\Omega}}$ for each model parameter estimated
   for the last 13 cycles.  The solar cycle with $\mb{\Omega} = \bar{\mb{\Omega}}$ 
   is the mean solar cycle.}     
   \label{table2}
\end{table}

\begin{table}[ht!]
   \centering
   \begin{tabular}{c c c c c c c c} 
  &  $a $ & $b$ & $c$ & $\kappa$ & $\beta_0$ & $\beta_1$ & $\beta_2$ \\
   \hline
 $a$ & 1.0000 & -0.5268 & -0.0243 & -0.4671 & 0.2148 & -0.4035 & -0.1543 \\
 $b$ &  & 1.0000 & -0.5270 & -0.0842 & -0.3877 & -0.1800 & -0.0196 \\
 $c$ &  &  & 1.0000 & 0.1596 & 0.6627 & 0.1721 & 0.2069 \\
 $\kappa$ &  &  &  & 1.0000 & -0.0022 & 0.8945 & 0.0733 \\
 $\beta_0$ &  &  &  &  & 1.0000 & -0.0921 & 0.1380 \\
 $\beta_1$ &  &  &  &  &  & 1.0000 & -0.1706 \\
 $\beta_2$ &  &  &  &  &  &  & 1.0000 
   \end{tabular}
   \caption{Correlation matrix for the model parameters estimated for the 
   previous 13 solar cycles 1850--2010.}
   \label{tablecorr}
\end{table}

\begin{table}[ht!]
   \centering
   \begin{tabular}{ c c c c c c c} 
   $\hat{a} $ & $\hat{b}$ & $\hat{c}$ & $\hat{\kappa}$ & 
   $\hat{\beta}_0$ & $\hat{\beta}_1$ & $\hat{\beta}_2$ \\
   {\scriptsize $[\mbox{day}^{-3}]$}   & {\scriptsize $[\mbox{day}]$} 
   & & {\scriptsize$[\mbox{day}^{-1}]$}  & {\scriptsize$[\mbox{day}^{-1}]$} 
     & {\scriptsize $[\mbox{day}^{-1}]$} & {\scriptsize $[\mbox{day}^{-1}]$} \\
   \hline
   13.23$\times 10^{-8}$ & 1401 & 0.6929 & 0.0766 & 18.74 & 1.891 & 
   3.751$\times 10^{-7}$
   \end{tabular}
   \caption{Bayesian parameter estimates for solar cycle 19 using 
   the mean cycle as a prior and including data for the entire cycle.}
   \label{table3}
\end{table}

\begin{table}[ht!]
   \centering
   \begin{tabular}{ c c c c c c c} 
   $\hat{a} $ & $\hat{b}$ & $\hat{c}$ & $\hat{\kappa}$ & 
   $\hat{\beta}_0$ & $\hat{\beta}_1$ & $\hat{\beta}_2$ \\
      {\scriptsize $[\mbox{day}^{-3}]$}   & {\scriptsize $[\mbox{day}]$} 
   & & {\scriptsize$[\mbox{day}^{-1}]$}  & {\scriptsize$[\mbox{day}^{-1}]$} 
     & {\scriptsize $[\mbox{day}^{-1}]$} & {\scriptsize $[\mbox{day}^{-1}]$} \\
   \hline
   8.2605$\times 10^{-8}$ & 1400.5 & 0.8894 & 0.05624 & 40.219 & 1.0995 & 1.621$\times 10^{-3}$
   \end{tabular}
   \caption{Bayesian parameter estimates for solar cycle 20 using the mean cycle as a prior
   and including data for the entire cycle.}
   \label{table4}
\end{table} 

\begin{table}[ht!]
   \centering
   \begin{tabular}{ c c c c c c c} 
   $\hat{a} $ & $\hat{b}$ & $\hat{c}$ & $\hat{\kappa}$ & 
   $\hat{\beta}_0$ & $\hat{\beta}_1$ & $\hat{\beta}_2$ \\
      {\scriptsize $[\mbox{day}^{-3}]$}   & {\scriptsize $[\mbox{day}]$} 
   & & {\scriptsize$[\mbox{day}^{-1}]$}  & {\scriptsize$[\mbox{day}^{-1}]$} 
     & {\scriptsize $[\mbox{day}^{-1}]$} & {\scriptsize $[\mbox{day}^{-1}]$} \\
   \hline
   4.2962$\times 10^{-8}$ & 1400.0 & 0.7804 & 0.09514 & 10.487 & 1.6496 & 0.0040
   \end{tabular}
   \caption{Bayesian parameter estimates for solar cycle 24, using the 
   mean cycle as a prior and including all sunspot data from 1 January 
   2009 to 31 March 2011.}
   \label{table5}
\end{table}
\end{article} 
\end{document}